**Continued activity of the 25th cycle: largest in 20 years. Ground-level enhancement and Forbush decrease**

B.Sargsyan and A.Chilingarian

A.I.Alikanyan National Lab (Yerevan Physics Institute)
Alikhanyan Brothers 2, Yerevan 36, Armenia, AM00036

## Abstract

After a very calm 24th solar activity cycle, the 25th cycle has already seen several interesting events. A Ground Level Enhancement (GLE 77) was observed on 11 November 2025 following an X5.1-class solar flare. A strong Forbush decrease (FD) occurred on 19–20 January 2026 during one of the most intense geomagnetic storms of Solar Cycle 25. Events were recorded coherently by the global neutron monitor (NM) network and by SEVAN detectors at multiple altitudes. Using SEVAN's light-spectrometric capabilities, we reconstruct energy-dependent spectra of "missing" neutrons and muons during the FD and compare them with corresponding spectra measured during GLE 77. The analysis demonstrates that FD and GLE signatures are intrinsically asymmetric: FDs selectively suppress the pre-existing galactic cosmic-ray population, whereas GLEs introduce an additional, harder particle component. Neutron and muon channels exhibit markedly different spectral behavior, particularly at higher deposited energies, reflecting their sensitivity to different primary-energy ranges. These results show that combined NM–SEVAN observations provide robust, complementary diagnostics of rigidity-dependent cosmic-ray modulation during extreme heliospheric disturbances.

## Key Points

• The Forbush decrease of 19–20 January 2026 was coherently registered by the global neutron monitor network and by SEVAN detectors, confirming strong rigidity-dependent suppression of galactic cosmic rays during an extreme geomagnetic storm.

• Energy-resolved SEVAN spectra show that neutron deficits are systematically larger than muon deficits, demonstrating that low-rigidity primaries are more strongly modulated than the higher-energy primaries responsible for ground-level muons.

• Comparison of FD and GLE spectra reveals a clear asymmetry between cosmic-ray suppression and enhancement, with distinct neutron–muon spectral behavior providing complementary diagnostics of heliospheric modulation.

## Keywords

## 1. Introduction

Ground Level Enhancements (GLEs) are high-energy events in which solar protons reach energies sufficient to generate secondary cascades detectable by ground-based detectors (Shea & Smart, 2012). GLEs are the highest-energy manifestation of solar energetic protons (SEP; Reames, 2013); more than 77 officially registered GLEs have been observed predominantly associated with major eruptive flares and CME-driven shocks (Bütikofer & Flückiger, 2015).

The particle fluxes measured at Earth's surface exhibit depletions (called Forbush Decreases - FDs) due to disturbances in near-Earth magnetic structures in response to propagated shocks and ICMEs (Maricic et al., 2014). FDs are the most frequent and easily detected phenomenon of solar modulation of galactic cosmic rays. Historically, more than 80 years ago, Scott Forbush was the first to relate these depletions in the cosmic radiation (CR) flux to solar eruptions (Forbush 1954).

FDs and GLEs are two distinct manifestations of solar–terrestrial interaction. FDs arise from the suppression of galactic cosmic rays (GCRs) by interplanetary magnetic structures associated with coronal mass ejections, whereas GLEs result from the injection of high-energy solar protons into the near-Earth environment. Although both phenomena are routinely observed by neutron monitors, their energy-dependent characteristics and their manifestations in different secondary particle channels remain insufficiently explored.

The SEVAN detector concept (Chilingarian et al., 2014), which combines neutron- and muon-sensitive channels at a single location, provides a more differential view of cosmic-ray modulation. While neutron monitors are primarily sensitive to lower-rigidity primaries, muons probe substantially higher primary energies. This work exploits simultaneous NM and SEVAN observations of the November 11, 2025, GLE and the January 2026 FD to reconstruct energy-dependent "missing-particle" spectra and to compare them with spectra measured during a GLE, focusing on meaningful spectral differences rather than on acceleration modeling.

## 2. GLE 77 on 11 November 2025, the second-largest in 20 years after GLE 69 on 20 January 2005

The Ground Level Enhancement (GLE) 77 occurred on 11 November 2025 in association with an X5.1-class solar flare, which reached its soft X-ray maximum at 10:04 UT, as recorded by GOES X-ray sensors. High-energy proton channels on GOES-18 showed a rapid intensity increase within several minutes after the flare peak, indicating prompt acceleration and near-scatter-free propagation of relativistic solar protons to 1 AU. Neutron monitors (NMs) and the SEVAN network detected the onset of GLE 77 at 10:12–10:14 UT (Chilingarian et al., 2025a). The peak response varied strongly among stations, reflecting their differing geomagnetic cutoff rigidities. In contrast to large polar events, the response at mid-latitude stations was modest, indicating a limited flux of very high-rigidity particles.

Figure 1 presents a time series of neutron flux enhancements, expressed as standard deviations (σ) relative to the pre-event baseline.

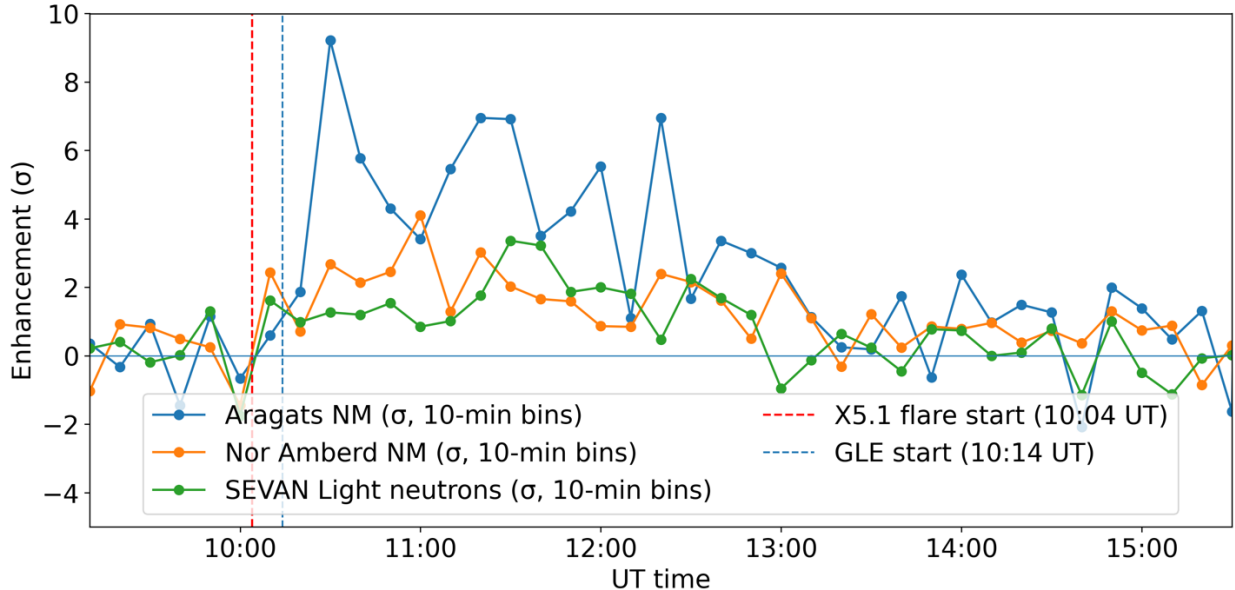

**Figure 1. Ten-minute time series of neutron count enhancements expressed in units of standard deviation (σ) for the Aragats Neutron Monitor, Nor Amberd Neutron Monitor, and SEVAN Light neutron detector during GLE 77 on 11 November 2025. The mean count rate and variance were calculated over the quiet interval 09:10–10:00 UT. Vertical dashed lines indicate the start of the X5.1 solar flare at 10:04 UT and the onset of the ground-level enhancement at 10:14 UT.**

Aragats neutron monitor (ARNM, 3200 m asl) exhibits significantly larger σ-amplitudes than the Nor Amberd neutron monitor (NANM, 2000 m asl), and the fine temporal structure consists of multiple peaks that are not strictly synchronized between the two stations.

The larger σ-amplitudes observed at ARNM are primarily a consequence of site altitude and atmospheric depth. ARNM operates at a higher altitude on Mt. Aragats; as a result, the secondary nucleonic cascade initiated by relativistic solar protons is less attenuated before reaching the detector. For a given primary proton rigidity, this leads to a higher neutron yield and a stronger relative enhancement at Aragats. When expressed in σ units, the effect is further amplified because the higher-altitude station typically exhibits lower relative statistical noise, increasing the signal-to-noise ratio for transient enhancements.

In addition, small differences in effective geomagnetic cutoff rigidity and asymptotic viewing directions between the two sites modify their sensitivity to anisotropic solar proton fluxes during the early phase of the event. Even for stations separated by only tens of kilometers, differences in geomagnetic transmission can translate into appreciable differences in the detected flux of near-threshold particles during a weak or moderate GLE.

The multiple, non-coincident peaks in the ARNM and NANM time series reflect a combination of anisotropic particle arrival and energy-dependent transport effects. During the early phase of GLEs, relativistic protons often arrive in narrow pitch-angle distributions. As the interplanetary magnetic field connection to the acceleration region evolves, different rigidity ranges may dominate the response at different times. Furthermore, local atmospheric conditions and the intrinsic stochasticity of secondary cascade development can modulate the neutron count rate on short timescales, contributing to the observed lack of perfect peak-to-peak correspondence. The fact that both stations nevertheless show a coherent onset at 10:12–10:14 UT confirms the solar origin of the event, while the differing fine structure highlights the sensitivity of mid-latitude neutron monitors to subtle variations in the relativistic solar proton population.

Overall, the stronger, more structured response of ARNM underscores the critical role of high-altitude neutron monitors in resolving weak GLEs, while the comparison with NANM demonstrates that even closely spaced stations can provide complementary information on particle anisotropy and energy-dependent transport during the earliest phases of solar energetic particle events.

The response of the SEVAN Light detector shown in Fig. 1 provides an important independent validation of GLE 77 at Aragats. SEVAN Light employs a 0.25 m² plastic scintillator optimized for neutron detection and equipped with an active veto layer, which efficiently suppresses the charged-particle component of the secondary cosmic-ray flux (see details of instrumentation in Chilingarian et al., 2025a). As a result, the registered signal is dominated by atmospheric neutrons produced in hadronic cascades initiated by relativistic solar protons.

Compared to neutron monitors, the absolute count rate of SEVAN Light is substantially lower, which naturally limits the achievable statistical significance when the data are expressed in σ units. Nevertheless, the detector exhibits a clear and temporally consistent enhancement starting at 10:12–10:14 UT, coincident with the GLE onset identified by both Aragats and Nor Amberd neutron monitors.

Previous studies of the 20 January 2005 GLE 69, based on high-energy muon data from Aragats, indicated a very hard solar-proton spectrum extending beyond 20 GeV (Bostanjan et al., 2007; Chilingarian, 2009) and developed methods to estimate GLE hardness from ground-based observations (Zazyan et al., 2011). This work directly builds on those findings by applying advanced spectrometric and coincidence techniques to a new GLE observed two decades later, highlighting both the rarity of such events and the ongoing nature of the research program. Analyzing this recent GLE provides a unique chance to revisit and significantly expand previous conclusions with improved experimental methods and to explore the implications of hard, high-energy solar-proton spectra for rapid space-weather prediction.

### 3. Forbush decrease on 19-20 January 2026, registered by Neutron Monitor and SEVAN networks

A long-duration X1.95 solar flare peaked at 18:09 UT on 18 January 2026 and originated from active region AR 4341, which at the time was near the central meridian of the Earth-facing solar disk. The flare was accompanied by a large coronal mass ejection, whose full-halo appearance in coronagraph imagery confirmed a substantial Earth-directed component. Although the bulk of the ejecta propagated eastward, the halo morphology indicated that a significant fraction of the plasma cloud was aligned along the Sun–Earth line, rendering the event highly geoeffective.

The interplanetary disturbance associated with this CME reached Earth on 19 January and triggered a major geomagnetic storm. From 19 to 21 January, geomagnetic activity reached severe storm levels (G4) for approximately 15 hours and remained at strong storm levels (G3) for more than 18 hours. This prolonged period of enhanced geomagnetic activity produced widespread auroral displays across most of Europe and the U.S. states, underscoring the event's exceptional strength and global impact. In terms of both intensity and duration, this storm ranks among the largest space-weather disturbances of Solar Cycles 23-25.

High-resolution geomagnetic measurements at Aragats revealed a rapid, pronounced depression in the X-component of the magnetic field, with a well-defined minimum at 21:26 UT on 19 January. The magnetic response exhibits a step-like decrease, consistent with sudden magnetospheric compression followed by enhanced ring-current development. The timing of this magnetic minimum provides a robust reference point for the arrival of the CME-driven shock and sheath at Earth.

Immediately following the geomagnetic response, a coherent Forbush decrease in the galactic cosmic-ray flux was observed, as recorded by multiple neutron monitors and the SEVAN network (see Figure 2). The onset of the decrease occurred within a narrow time window but showed a modest station-to-station dispersion, reflecting differences in geomagnetic cutoff rigidity and asymptotic viewing directions. Nor Amberd registered the earliest onset, followed closely by Lomnický Štít, Oulu, and Apatity. The depth of the Forbush decrease reached approximately 15–20% below the pre-event level, indicating efficient suppression of galactic cosmic rays by the enhanced magnetic field and turbulence associated with the CME sheath and magnetic cloud.

A corresponding suppression was detected by the SEVAN network at middle-altitude stations, including Aragats, Lomnický Štít, and Musala. In the configuration used here, the SEVAN "010" channel (signal in the middle scintillator only) is dominated by secondary neutrons with a limited admixture of muons, and therefore provides a response function that is closely comparable to neutron-monitor observations. The time profiles of the disturbances recorded by SEVAN closely reproduce the canonical Forbush-decrease morphology seen in the neutron monitors: an abrupt onset, a rapid transition to a depressed level, and a slower recovery. This similarity in shape indicates that both detector systems are tracking the same large-scale modulation of the secondary particle field driven by the CME disturbance. The remaining differences are quantitative rather than qualitative. In particular, the SEVAN onset times can appear slightly more dispersed and, in some cases, modestly later than those inferred from the neutron monitors, primarily because the SEVAN "010" channel has different counting statistics and an energy–response mixture that is not identical to a standard neutron monitor

The temporal ordering of the observations is unambiguous. The geomagnetic disturbance, marked by the minimum of the X-component, precedes the onset of the Forbush decrease in neutron monitors, which in turn coincides with or slightly precedes the suppression observed by the SEVAN network. This sequence establishes a clear causal chain linking the CME-driven interplanetary shock to magnetospheric compression and subsequent cosmic-ray modulation. The consistency of timing and behavior across independent measurement techniques demonstrates the value of coordinated geomagnetic, neutron-monitor, and secondary-particle observations for diagnosing the evolution and impact of extreme space-weather events.

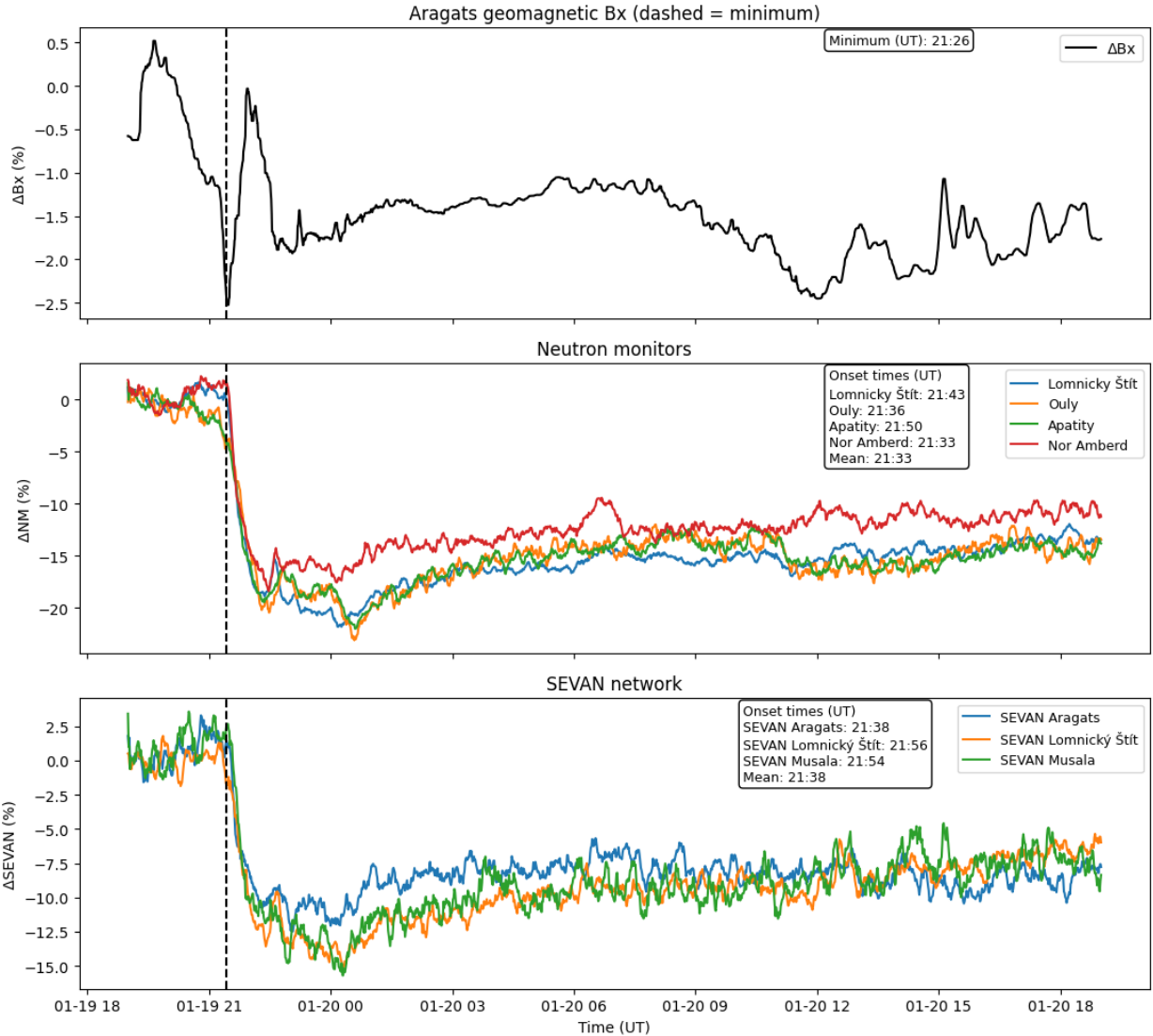

**Figure 2. Time-aligned response of the geomagnetic field, neutron monitors, and SEVAN detectors during the January 2026 Forbush decrease event. The top panel shows the filtered X-component of the geomagnetic field at Aragats, expressed as a percentage deviation from the pre-event baseline; the vertical dashed line marks the time of the magnetic field minimum associated with the arrival of the CME-driven disturbance. The middle panel presents relative count-rate variations recorded by four neutron monitors (Nor Amberd, Lomnický Štít, Oulu, and Apatity), with onset times indicated for each station. The bottom panel shows the corresponding suppression observed by the SEVAN network at Aragats, Lomnický Štít, and Musala. All data are normalized to the same baseline interval, illustrating the coherent temporal progression from geomagnetic disturbance to primary and secondary cosmic-ray modulation.**

In Figure 3, we compare the FD measured by the Aragats Solar Neutron Telescope (ASNT) with that measured by the NANM. The FD amplitude measured by the neutron monitor is much larger than that measured by ASNT (mostly muons). The spikes during the recovery phase are due to gamma-ray contamination. During FD, a major snowstorm hit Aragats, and a wind-driven gamma-ray storm occurred. Detector count rates are enhanced by gamma radiation from isotopes brought by strong winds, which pump them into the building where the detectors are located (see details in [1]).

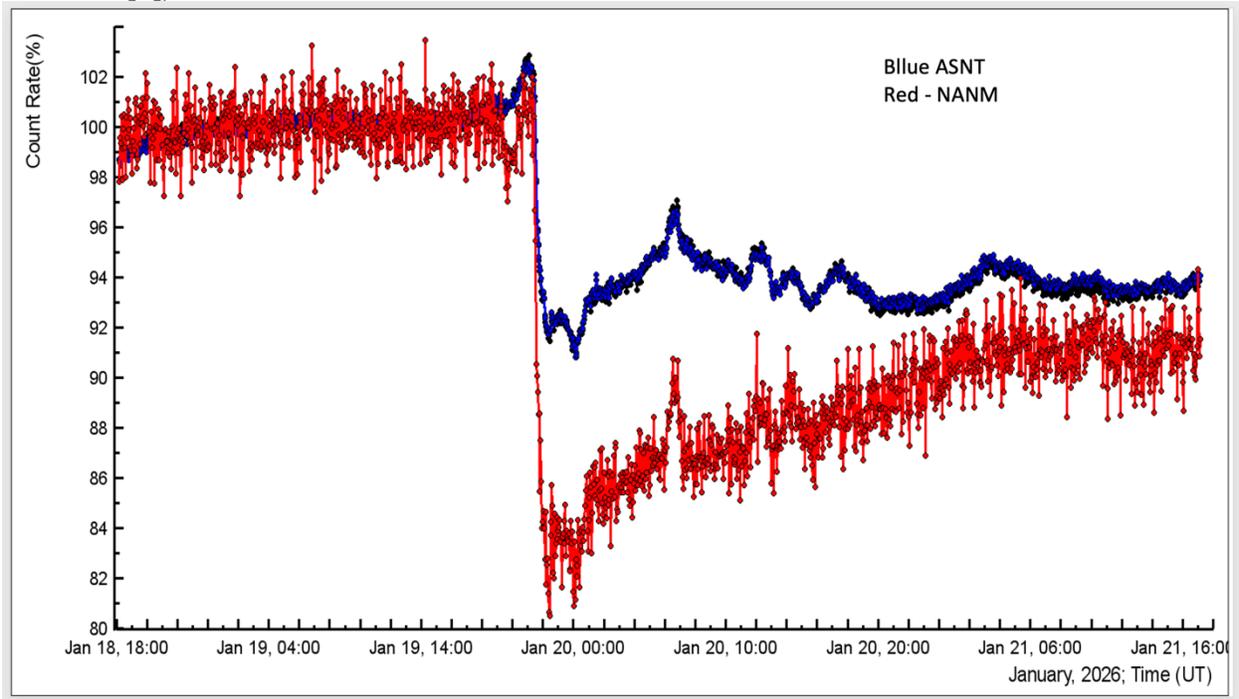

**Figure 3. Comparison of the Forbush decrease recorded by the ASNT and NANM. The neutron monitor exhibits a substantially larger FD amplitude than the muon-dominated ASNT, reflecting the stronger modulation of lower-energy primary cosmic rays. The temporal evolution of the disturbance is similar for neutrons and muons, indicating a common interplanetary driver. Spikes during the recovery phase are caused by gamma-ray contamination associated with a wind-driven gamma-ray storm during a major snowfall at Aragats (Chilingarian et al., 2025b).**

The Forbush decrease recorded by the Aragats Solar Neutron Telescope (ASNT) and the Nor Amberd Neutron Monitor (NANM). In the presented configuration (without veto on charged particles), the ASNT is dominated by muons. The markedly different FD amplitudes in neutron and muon channels follow naturally from the rigidity dependence of heliospheric modulation during CME-driven disturbances. Neutron monitors respond primarily to the nucleonic component generated by primaries with characteristic rigidities of a few to roughly 10–20 GV, depending on geomagnetic cutoff and atmospheric depth, whereas ground-level muons are preferentially produced by higher-rigidity primaries, typically tens of GV and above. Because the fractional suppression of galactic cosmic rays decreases with increasing rigidity, the neutron-monitor decrease is expected to be larger than the muon decrease. The present event exhibits exactly this behavior: the neutron channels show a deep suppression (order 15–20%), while the muon-dominated SEVAN response is noticeably smaller. Importantly, the similarity of onset and recovery shapes across the two particle populations indicates that the underlying interplanetary driver is common; the amplitude contrast is therefore interpreted as an energy–rigidity effect rather than a difference in the physical origin of the disturbance.

In Figure 4, we present the famous Halloween events of 2003. The amplitude of FD reaches 23% at Aragats. However, it is a two-stage FD; the second CME arrives before the recovery face and deepens the FD.

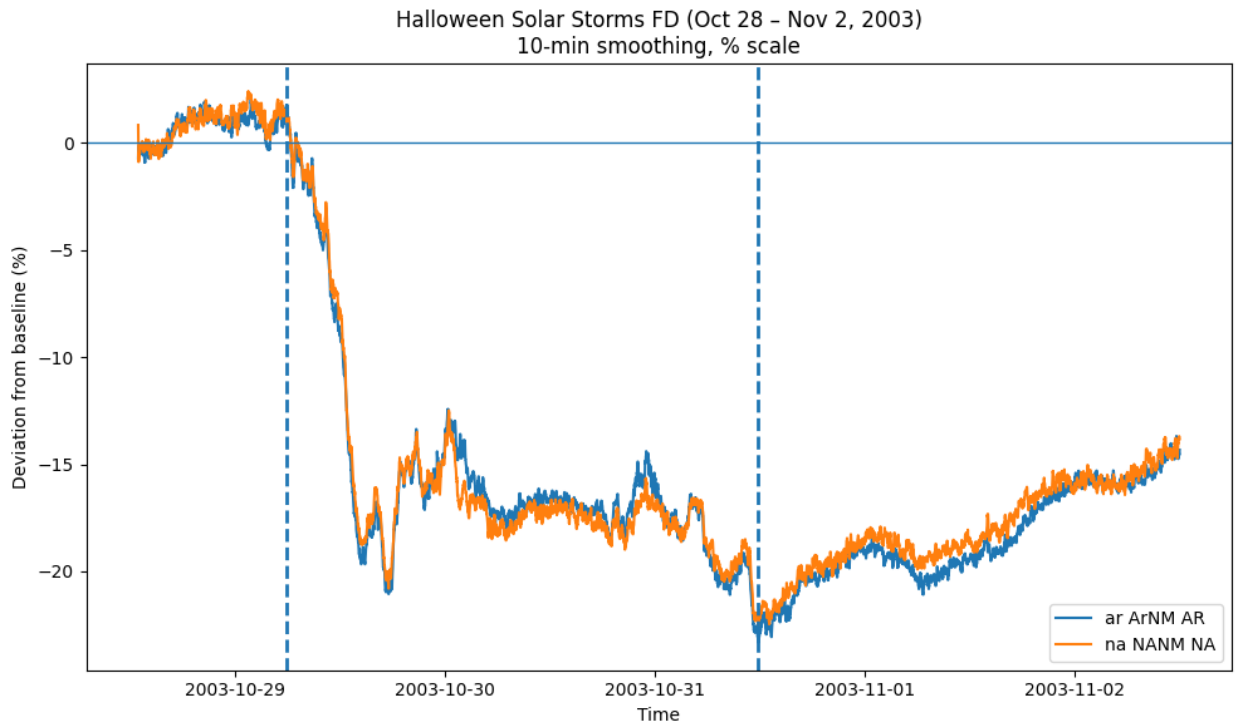

**Figure 4. The Forbush decrease during the Halloween events of October–November 2003, shown for comparison with the January 2026 event. The FD amplitude at Aragats reached about 23%, the disturbance had a two-stage structure, with a second CME arriving before recovery from the first. Although the January 2026 FD was driven by a single CME and reached only a slightly smaller amplitude, its duration and global impact make it comparable in significance to the Halloween events.**

When this structural difference is taken into account, the January 2026 event is comparable in overall significance to the Halloween storms. The duration of severe and strong geomagnetic conditions, the global extent of auroral activity, and the coherent response across geomagnetic, neutron-monitor, and SEVAN observations place the January 2026 disturbance among the most important space-weather events of the current solar cycle.

## 4. Energy spectra of "missing particles", comparison with GLE 77

The capability of SEVAN Light detectors to reconstruct energy-release spectra of secondary neutrons and muons was originally developed and successfully demonstrated for Ground Level Enhancements, in particular for GLE 77, where the simultaneous observation of neutrons and muons at a single location provided direct information on the spectral hardness of relativistic solar particles, allowing for properties of the solar accelerator.

For Forbush decreases, the same instrumental capability can be applied in an inverted, yet physically meaningful, way. Instead of excess particles, Forbush decreases are characterized by a *deficit* of cosmic rays relative to a quiet-time reference state preceding the interplanetary disturbance. Using SEVAN Light data, one can define "missing particle spectra" as the difference between the quiet-time secondary particle spectra and those measured during the main phase of the Forbush decrease. These spectra quantify, as a function of secondary energy, which particles are removed from the flux during the ICME-magnetosphere interaction.

The concept of missing particle spectra can be particularly useful because it preserves information on energy dependence that is lost in integral count-rate measurements. While neutron monitors provide high-precision records of intensity reductions, their response integrates over a broad range of primary rigidities, making it difficult to disentangle whether the suppression is dominated by low-rigidity particles or extends uniformly across higher rigidities, and how it evolves with time. The simultaneous reconstruction of neutron and muon deficit spectra potentially enables a comparative analysis of "missing" particle populations originating from different primary rigidity ranges.

Neutrons predominantly reflect the modulation of lower-rigidity primary cosmic rays, whereas muons probe higher rigidities due to the kinematics of pion production and decay in the atmosphere. Consequently, differences in the depth, spectral shape, and temporal evolution of neutron and muon deficits directly indicate the rigidity dependence of the Forbush decrease. A stronger suppression of neutron spectra relative to muons implies enhanced exclusion of lower-rigidity particles, typically associated with strong magnetic turbulence in the ICME sheath. Conversely, comparable suppression across both components suggests a broader rigidity cutoff, consistent with coherent magnetic cloud structures and large-scale field draping.

The relationship between missing-particle spectra and magnetospheric disturbances becomes especially relevant during intense geomagnetic storms, such as the January 19–21, 2026 event, when prolonged G4–G3 conditions indicate strong coupling between the interplanetary magnetic field and the terrestrial magnetosphere. During such periods, magnetospheric compression, variations in cutoff rigidity, and enhanced geomagnetic activity can further modify particle access to the atmosphere. The energy-dependent deficits observed in neutron and muon spectra therefore reflect not only heliospheric modulation by the ICME but also the magnetosphere's response to extreme external forcing.

In this framework, missing-particle spectra provide a quantitative diagnostic of how different rigidity ranges of cosmic rays are suppressed during combined interplanetary and magnetospheric disturbances. Geomagnetic storm severity is formally defined by geomagnetic indices such as Kp, Ap, and Dst, which quantify disturbances of the geomagnetic field and are independent of particle flux measurements. Accordingly, SEVAN Light observations do not provide a measure of storm intensity in the geomagnetic sense. Instead, the value of neutron and muon missing-particle spectra lies in their ability to characterize the rigidity-dependent modulation of cosmic rays produced by interplanetary disturbances and their coupling to the magnetosphere. When interpreted alongside standard geomagnetic indices, these spectral deficits provide complementary insights into how ICME-driven magnetic structures modify particle access to the atmosphere during geomagnetic storms, without replacing or redefining established storm severity metrics.

Figure 5 shows the systematic deficit in neutron and muon spectra during geomagnetic storms relative to undisturbed magnetospheric conditions. To emphasize the relative deficit in muons and neutrons (the quantity measured by the surface detectors), we normalized the integral spectra to the 10 MeV intensities. The figure reflects the temporary suppression of galactic cosmic rays (GCRs) by interplanetary magnetic structures. The suppression is energy dependent, with a stronger effect at lower rigidities and a gradual weakening of the neutron spectrum toward higher energies. The muon spectrum is less affected than the neutron spectrum, consistent with the fact that muons are produced predominantly by higher-energy primaries that are less sensitive to GCR modulation.

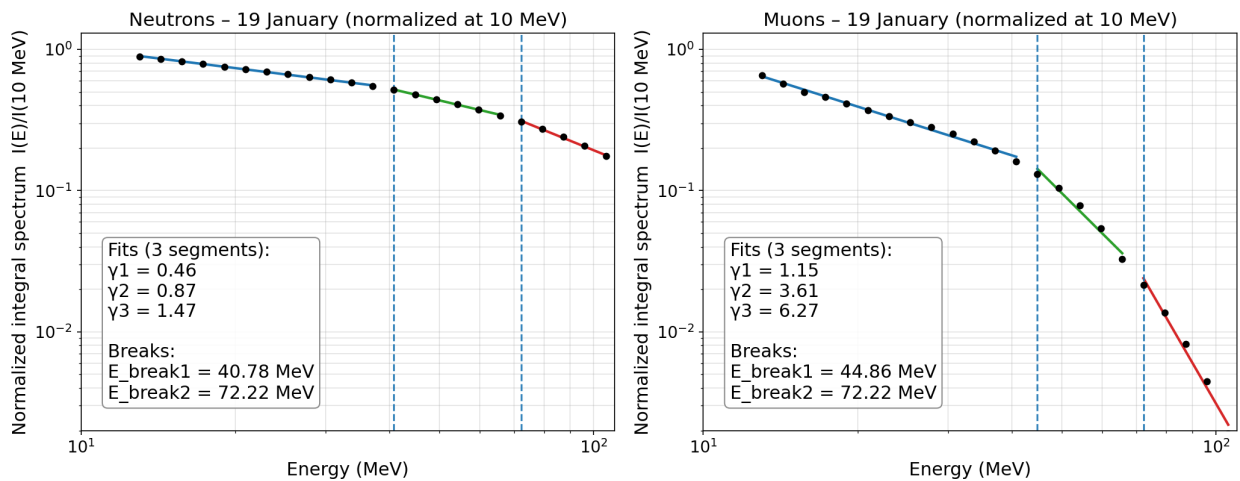

**Figure 5. Relative integral energy spectra of "missing" neutrons and muons during FD and GLE normalized to intensities at 10 MeV.**

The FD amplitude is larger for neutrons than for muons at all energies, in agreement with Fig. 3. Neutrons detected at ground level are produced predominantly by lower-energy primary protons near the geomagnetic cutoff and within the rigidity range most strongly modulated by interplanetary magnetic structures. Muons are produced by higher-energy primaries because pion production and decay require substantially higher primary energies and deeper shower development. Because Forbush decreases suppress the galactic cosmic-ray (GCR) flux most efficiently at low rigidities, the neutron component experiences a stronger reduction, whereas the muon component—fed by higher-rigidity primaries—is less affected.

The neutron deficit spectrum shows a relatively hard low-energy segment, followed by a moderate steepening above the first break and a stronger steepening at higher energies. This structure is consistent with a mixed contribution from primaries near the cutoff rigidity, competition between hadronic production and atmospheric attenuation, and a progressive weakening of FD influence with increasing primary rigidity.

In contrast, the muon spectrum fitted indices are steeper across all segments, with the highest segment steepening much more rapidly than for neutrons (6.27 versus 1.47). This softness reflects the fact that muons sample a narrower, higher-energy primary population, where FD modulation is already reduced.

Figure 6 compares the integral spectra obtained from SEVAN Light during the GLE and during the Forbush-decrease deficit ("missing" particles), shown separately for neutrons and muons. The key takeaway is that neutrons and muons respond very differently at the high-energy end of the spectrum. In both panels, the GLE enhancement remains substantial at higher deposited energies, whereas the FD deficit steepens and becomes relatively weaker. This divergence is markedly stronger in the muon channel than in the neutron channel. In practical terms, the high-energy tail is far more robust for muons in the GLE than for the FD deficit, consistent with the fact that ground-level muons preferentially sample higher-energy primaries and are therefore sensitive to the presence or absence of a multi-GeV component, whereas the neutron response is fed efficiently by a broader and generally lower primary-energy range.

A second, nontrivial feature is the local structure in the muon spectrum near 40–50 MeV. The presence of a "bump" (a local hardening/flattening relative to adjacent energies) indicates that the muon channel is not described well by a single smooth power-law-like behavior over the full range. A plausible interpretation, consistent with cascade physics and detector selection, is that this energy interval corresponds to a transition region where the muon sample recorded by the given channel changes composition and/or production conditions: the relative contribution of more penetrating muons produced higher in the atmosphere (and originating from higher-energy primaries) increases compared with lower-energy secondaries and/or mixed-event backgrounds. In other words, the muon channel exhibits a transition in the effective parent population around this deposited-energy band, which can appear as a local excess in an integral spectrum when two components with different slopes overlap. This feature is much less pronounced in the neutron spectrum, again emphasizing that the muon channel carries additional spectral-structure information at higher energies that is not mirrored in the neutron response.

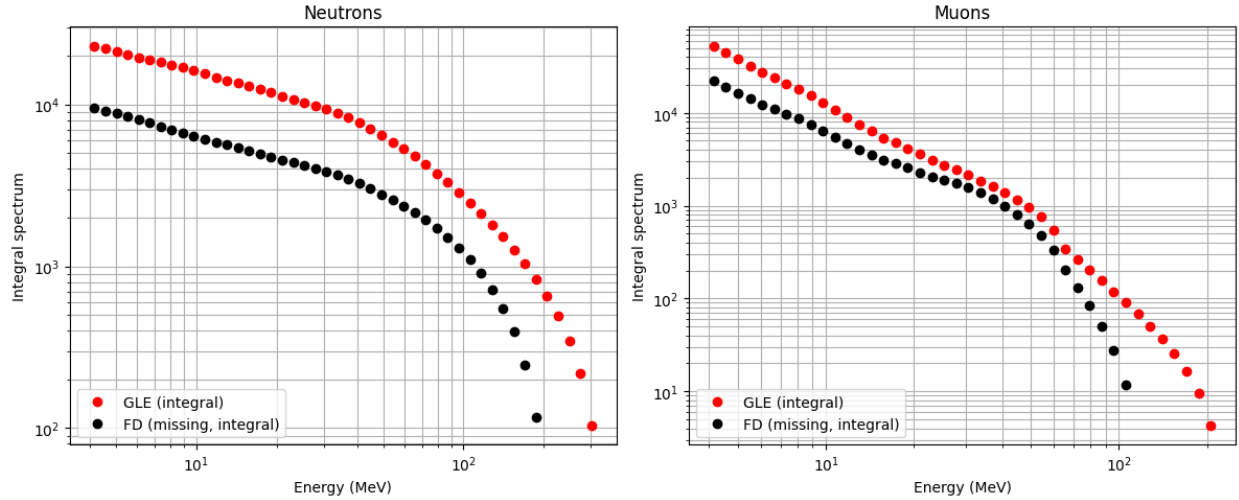

**Figure 6. Integral energy spectra of neutrons (left) and muons (right) during the GLE and the Forbush decrease (FD), plotted together starting from 10 MeV. The FD spectra represent the energy-dependent deficit ("missing" particles) relative to background GCR conditions, while the GLE spectra show the enhancement due to solar energetic particles. The comparison illustrates the intrinsic asymmetry between suppression and injection processes, as well as the different responses of the neutron and muon channels.**

## 5. Conclusion

The Forbush decrease of 19–20 January 2026 was unambiguously detected by both the global neutron monitor network and the SEVAN detector system, demonstrating the coherence of cosmic-ray modulation across independent instruments and particle populations. Neutron monitors recorded a deep suppression, reflecting efficient exclusion of low-rigidity galactic cosmic rays, while the muon-dominated SEVAN response showed a significantly smaller amplitude decrease, consistent with the higher primary energies required for muon production.

Energy-resolved "missing-particle" spectra show that the FD amplitude is systematically larger for neutrons than for muons at all measured energies. This behavior directly reflects the rigidity dependence of heliospheric modulation: particles near the geomagnetic cutoff are suppressed most efficiently, whereas higher-rigidity primaries feeding the muon channel are less affected. The neutron spectra steepen only moderately with energy, whereas the muon deficit spectra are substantially softer, indicating that the FD influence rapidly diminishes at higher energies. Comparison with GLE spectra highlights the fundamental asymmetry between suppression and enhancement processes. GLE spectra retain a pronounced high-energy component, especially in muons, whereas FD deficit spectra steepen strongly and fade toward higher energies. A notable feature is a localized structure in the muon spectra around 40–50 MeV deposited energy, suggesting a transition in the effective parent population or in the atmospheric cascade contribution, which is absent in the neutron channel. This feature emphasizes that muon measurements carry additional spectral information that cannot be inferred from neutron data alone.

Overall, the joint use of neutron monitors and SEVAN spectrometry provides a powerful, physics-driven approach to diagnosing rigidity-dependent cosmic-ray modulation during extreme solar events. While neutron monitors remain the primary tool for detecting and quantifying FDs, SEVAN measurements add essential spectral discrimination, enabling separation of low- and high-energy responses. This combined strategy is particularly valuable for characterizing severe space weather events in Solar Cycle 25 and beyond.

## Acknowledgements

The authors acknowledge the support of the Science Committee of the Republic of Armenia (Research Project No. 21AG-1C012).

## Data Availability Statement

The neutron and muon data from the ASNT and the SEVAN Light detector that support the findings of this study are available from the corresponding author upon reasonable request. Processed data products and reconstructed spectra used in this analysis are derived from these measurements using standard procedures described in the manuscript. GOES proton data are publicly available from NOAA archives.

## Conflict of Interest

The authors declare no conflicts of interest relevant to this study.